\documentclass[5p,twocolumn,times,number]{elsarticle}
\usepackage{lineno}
\usepackage{graphicx}
\usepackage{amsmath}   
\usepackage{epsfig}
\begin{document}
\begin{frontmatter}
\title{Longitudinal uniformity, time performances and irradiation test of pure CsI crystals}

\author[add1]{M.~Angelucci}
\author[add2]{O.~Atanova}
\author[add3]{S.~Baccaro}
\author[add3]{A.~Cemmi}
\author[add1]{M.~Cordelli}
\author[add1,add4]{R.~Donghia\corref{cor}}\ead{raffaella.donghia@lnf.infn.it}
\author[add1]{S.~Giovannella}
\author[add1]{F.~Happacher}
\author[add1]{S.~Miscetti}
\author[add1]{I.~Sarra}
\author[add1]{S.R.~Soleti}
\cortext[cor]{Corresponding author}
\address[add1]{Laboratori Nazionali di Frascati dell'INFN, Frascati, Italy}
\address[add2]{Joint Institute for Nuclear Research, Dubna, Russia}
\address[add3]{ENEA UTTMAT-IRR, Casaccia R.C., Roma, Italy}
\address[add4]{Dipartimento di Fisica, Universit\'a Roma Tre, Roma, Italy}
\begin{abstract}
To study an alternative to BaF$_2$, as the crystal choice for the Mu2e calorimeter, thirteen pure CsI crystals from Opto Materials and ISMA producers have been characterized by determining their light yield (LY) and longitudinal response uniformity (LRU), when read with a UV extended PMT. The crystals show a LY of $\sim$ 100 p.e./MeV ($\sim$ 150 p.e./MeV) when wrapped with Tyvek and coupled to the PMT without (with) optical grease. The LRU  is well represented by a linear slope that is on average $\delta \sim $ -0.6 \% $/$cm. The timing performances of the Opto Materials crystal, read with a UV extended MPPC, have been evaluated with minimum ionizing particles. A timing resolution of $\sim$ 330 ps ($\sim$ 440 ps) is achieved when connecting the photosensor to the MPPC with (without) optical grease. The crystal radiation hardness to a ionization dose has also been studied for one pure CsI crystal from SICCAS. After exposing it to a dose of 900 Gy, a decrease of 33\% in the LY is observed while the LRU remains unchanged.
\end{abstract}
\end{frontmatter}
\section{Introduction}
The Mu2e calorimeter is designed to achieve an energy resolution of $\mathcal{O}$(5\%) and a time resolution of $\mathcal{O}$(500 ps) for 104.97 MeV electrons coming from the muon to electron conversion process in Aluminum. The baseline calorimeter design consists of two disks of BaF$_2$ scintillating crystals readout by a new generation of UV extended ``solar blind'' APDs \cite{TDR}.  Since these APDs are still in the development phase, single crystals of pure CsI have been tested as a backup alternative.
\section{Experimental Setup}
Thirteen pure CsI crystals have been tested: 2 from Opto Materials (Italy), of dimensions $3\times 3 \times 20$ cm$^3$, and 11 from ISMA (Ukraine), out of which 7 have the same dimension (short), while other 4 (long) are longer and have dimension of $2.9\times 2.9 \times 23$ cm$^3$. These crystals present a large improvement on longitudinal transmittance, with respect to a SICCAS crystal (China) used as reference, reaching  $\sim$ 80\%  at a wavelength of 350 nm.\\
To measure the light yield (LY) and longitudinal response uniformity (LRU) of each crystal, a $^{22}$Na source has been used to illuminate them in a few mm$^2$ region. The source is placed between the crystals and a small tagging system, constituted by a ($3\times 3\times 10$) mm$^3$ LYSO crystal, read by a ($3\times 3$) mm$^2$ MPPC. One of the two back-to-back 511 keV annihilation photons produced by the source, is tagged by this monitor while the second one is used to calibrate the crystal under test. The crystals were optically connected to a 2'' UV extended EMI PMT.\\
For each crystal, a longitudinal scan has been done in eight points, with 2 cm steps, with respect to the PMT. During the scan, the source and the tag were moved together, along the crystal axis.\\
To study the dependence of the response on the wrapping material, some crystals have been tested by wrapping them with different reflector materials: Tyvek, Teflon or aluminum. The wrapping foils cover both the four longitudinal surfaces and the side opposite to the PMT. The effect of the Bluesil Past-7 silicon grease has also been studied for some crystals in this sample. 
\begin{figure}[!h]
\centering
\epsfig{file=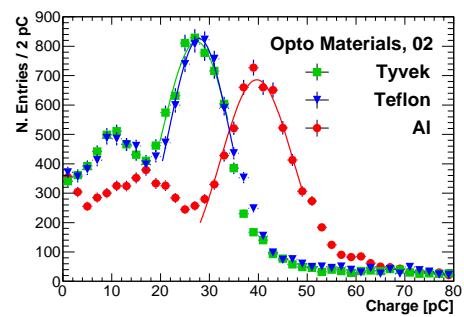,width=0.35\textwidth}
\caption{Charge distribution obtained with the source in the central scan position, when wrapping the crystals with different reflector materials. The gaussian fits are used to extract the 511 KeV peak position.}
\label{Fig:charge1}
\end{figure}
\section{Light Yield}
Pure CsI signals are typically within 300 ns from the start of the the acquisition window, with a full width of 150 ns. The charge ($Q$) is obtained integrating in the range ($0\div 300$) ns the amplitude of the signal. Finally the charge integral is corrected for calibration factors to be expressed in  pC. 
In Fig.~\ref{Fig:charge1}, the distribution of the charge for one of the crystals under test in the central scan position is shown for different wrapping materials. A clear peak due to the 511 keV photon is visible, only a cut on the relative timing between source and tag has been used. 
An asymmetric Gaussian fit is applied to extract the peak position ($\mu_{Q}$) corresponding to the annihilation photon and then the LY is evaluated as: $
\frac{Np.e.}{MeV} = \frac{\mu_{Q [pC]}}{G_{PMT}\cdot E_{\gamma} [MeV] \cdot q_{e^-} [pC]},
$
where $q_{e^-}$ is the electron  charge, $E_\gamma$ is the energy of the annihilation photon and $G_{PMT}$ is the PMT gain, 3.8$\times 10^6$, measured comparing to a calibrated SiPM. The wrapping with Teflon or Tyvek provides the best response: the LY increases of  about 20\% with respect to the aluminum configuration (from 80 p.e./MeV to 90-100 p.e./MeV).
Due to the fragility of Teflon, all other tests were performed by wrapping the crystals with two Tyvek layers of 100 $\mu$m each.
Measurements on the other crystals confirmed that the LY is about 90-130 p.e./MeV when coupling the crystal to the PMT without grease. Coupling with grease corresponds to an additional LY increase of  $\sim$  60 \%. 
\begin{figure}[!h]
\centering
\epsfig{file=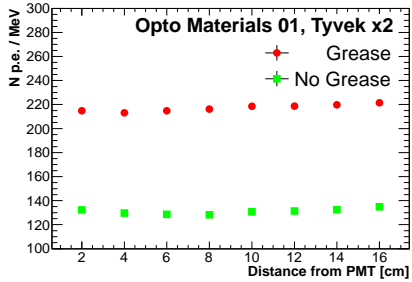,width=0.32\textwidth}
\caption{LY obtained for the Opto Materials N.1 with the source placed in 8 different positions with respect to the PMT.}
\label{Fig:LY}
\end{figure}
\section{Longitudinal Response Uniformity}
\begin{figure}[!h]
\centering
\epsfig{file=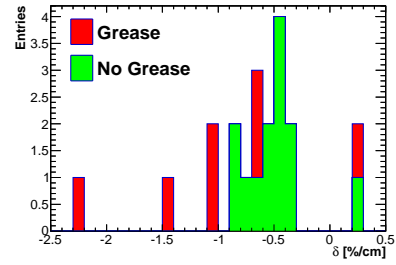,width=0.30\textwidth}
\caption{Slopes distribution (in \%/cm) provided by the linear fit on the LY normalized as function of the distance from the PMT.}
\label{Fig:slope}
\end{figure}

All crystals have been tested using just one orientation with respect to the PMT.  To evaluate the LRU, the LY has been normalized to the LY in the central position. This ratio is plotted as a function of the distance of the source from the PMT and fit with a linear function. The LRU is represented by the slope of the fit and shows an average of $\delta[\%/cm] \sim$ -0.6 \%/cm.
The slopes obtained are reported in the histogram of Fig.~\ref{Fig:slope}. A better LRU is reached without using grease in the optical contact.
\section{Timing Resolxution}
\begin{figure}[!h]
\centering
\epsfig{file=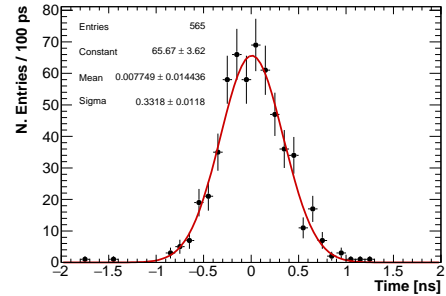,width=0.34\textwidth}
\caption{Time distribution (in ns)  of the Opto Material 01 crystal, wrapped with Tyvek and coupled with grease to the MPPC.}
\label{Fig:TR}
\end{figure}

To evaluate the timing performance one Opto Material crystal wrapped with Tyvek has been optically connected to a new UV extended MPPC array (16 3$\times$3 mm$^2$ cells with 50 $\mu$m pixels), both with and without optical grease.  The time response has been determined using minimum ionizing particles (MIP) from cosmic rays, selected by the coincidence  of 2 ``finger'' scintillators ($1\times1\times3$) cm$^3$, each readout by a fast  PMT, positioned one above and one below the crystal.\\
The signal waveforms were fit by a degree-4 polinomial function. A constant fraction method has been used to determine the crystal start time. To reduce the trigger time jitter, the half sum of the finger timing has been subtracted. A time resolution of $\sim$330 ps ($\sim$ 409 ps ), with (without) 
optical grease is achieved as shown in Fig.\ref{Fig:TR}. This result corresponds to an energy deposit of $\sim$ 22 MeV, that is the average energy deposited by a MIP in the crystal.
\section{Radiation hardness}
The SICCAS reference crystal  has also been exposed to a large ionization dose to test its radiation hardness. The test  has been carried out  at CALLIOPE (ENEA $\gamma$ irradiation facility \cite{Calliope}) where a $^{60}$Co source has been used to irradiate crystals up to 900 Gy.
After irradiation the LY of the crystal decreases by 33\%, as expected \cite{rad}. No changes were observed in the LRU. 


\begin{thebibliography}{10}
\expandafter\ifx\csname url\endcsname\relax
  \def\url#1{\texttt{#1}}\fi
\expandafter\ifx\csname urlprefix\endcsname\relax\def\urlprefix{URL}\fi
\bibitem{TDR}
L.~Bartoszek et al., {(Mu2e experiment)}, "Mu2e Technical Design Report"
	arXiv:1501.05241
\bibitem{Calliope}
S.~Baccaro et al., {Calliope $\gamma$ irradiation facility at ENEA-Casaccia} , TDR RT/2015/13/ENEA, ISSN/0393-3016
\bibitem{rad}
R.Y.~Zhu, "Precision crystal calorimetry in high energy physics"
	arXiv:hep-ex/9903023
\end{thebibliography}
\end{document}